\documentclass[prd,aps]{revtex4}
\usepackage{epsfig}
\def\be{\begin{eqnarray}}
\def\ee{\end{eqnarray}}
\def\beq{\begin{equation}}
\def\eeq{\end{equation}}
\def\ba{\begin{array}}
\def\ea{\end{array}}

\def\bec{\begin{center}}
\def\ec{\end{center}}

\begin{document}


\preprint{
\begin{tabular}{l}
\hbox to\hsize{\hfill KAIST-TH 2006/04}\\[-2mm]
\hbox to\hsize{\hfill hep-ph/yymmdd}\\[-3mm]
\hbox to\hsize{\hfill November 2006}\\[-3mm]
\end{tabular}
}

\title{
The minimal model of fermionic dark matter}

\author{Yeong Gyun Kim}
\email{ygkim@muon.kaist.ac.kr}

\affiliation{
ARCSEC, Sejong University, Seoul 143-747, Korea
}

\author{ Kang Young Lee }
\email{kylee@muon.kaist.ac.kr}
\vskip 1.0cm

\affiliation{
Department of Physics, Korea Advanced Institute of Science and Technology,
Daejeon 305-701, Korea
}

\date{\today}

\begin{abstract}

We explore the minimal extension of the Standard Model
with fermionic cold dark matter.
The interactions between the dark matter and the
Standard Model matters are described by the non-renormalizable
dimension-5 term.
We show that the measured relic abundance of the cold dark matter
can be explained in our model and predict the direct detection 
cross section.
The direct search of the dark matter provides severe constraints 
on the mass and coupling of the minimal fermionic dark matter
with respect to the Higgs boson mass.

\end{abstract}

\pacs{PACS numbers: 95.35.+d, 14.80.-j }

\maketitle


\section{Introduction}

Recently there has been a growing interest in the dark matter (DM)
of the universe as an interface between the cosmology and 
the particle physics, and between the astrophysical observation and 
the collider experiment.
The precise measurement of the relic abundance of the cold dark matter (CDM)
in the universe has been obtained from the Wilkinson microwave anisotropy probe (WMAP) 
data on the Cosmic Microwave Backgroun radiation
as \cite{WMAP}.
\be
0.085 < \Omega_{\rm CDM} h^2 < 0.119~(2\sigma~ \rm{level}),
\ee
where $\Omega$ is the energy density of the universe nomalized by the critical density
and $h \simeq 0.7$ is the scaled Hubble constant in units of 100 km/sec/Mpc.

Lots of CDM candidates have been suggested in various new physics
models beyond the Standard Model (SM).
In the supersymmetric model with conserved $R$-parity, 
the lightest supersymmetric particle (LSP) is stable 
and a good candidate of the CDM. \cite{SUSY,susydm}.
The KK-parity of the extra dimensional model suggests
a Kaluza-Klein dark matter \cite{KK}.
When the $T$-parity is conserved in the little Higgs model, 
the lightest $T$-odd particle, likely the heavy photon
can be the dark matter candidate \cite{littlehiggs}.

The minimal extension of the SM with the CDM is achieved
by introducing a singlet particle under the SM gauge group
for the new particle to be protected from the gauge interaction.
The model with a singlet scalar field with $Z_2$ parity
has been considered as the simplest candidate of the 
nonbaryonic cold dark matter
\cite{zee,burgess,murayama}.
Such a model contains three new parameters
and a neutral scalar degree of freedom which interacts with the SM matter
only via the coupling to the Higgs boson.
It is shown that the model with a singlet scalar can explain 
$\Omega_{\rm CDM} h^2$, and satisfies present experimental bounds 
for direct detection and collider signature.

In this work, we propose a model with a Dirac fermion 
which is a singlet under the SM gauge group
as a minimal model of fermionic dark matter.
We assign the baryon number and lepton number 
for the new fermion to be zero.
Our model does not need the additional symmetry like $Z_2$ parity,
$R$-parity, KK-parity nor $T$-parity
but require that the baryon number and lepton number be conserved
in order to avoid the mixing between the new fermion
and the SM fermions.
Then the dark matter can couple to only the Higgs boson
among the SM matter.
Since the Higgs field in the SM Lagrangian is SU(2) doublet,
the interaction is given by the dimension five term
${\cal L}_{\rm int} \sim -(1/\Lambda) |H|^2 \bar{\psi} \psi$
by introducing the new scale $\Lambda$,
where $H$ is the SM Higgs doublet and $\psi$ the dark matter fermion. 

This paper is organized as follows : In section 2, we review the model
by adding a singlet fermion to the SM.
The relic density of the CDM is evaluated in section 3
and the direct detection of the CDM is investigated in section 4.
Finally we conclude in section 5.

\section{The Model}

We introduce a singlet fermion $\psi$ under the SM gauge group.
The Lagrangian consists of
\be
{\cal L} = {\cal L}_{\rm SM} + {\cal L}_{\rm DM} + {\cal L}_{\rm int},
\ee
where the dark matter Lagrangian
\be
{\cal L}_{\rm DM} = \bar{\psi} ~i \gamma^\mu \partial_\mu \psi
            -m_0 \bar{\psi} \psi.
\ee
The leading interaction term between the dark matter and the SM fields 
is given by the dimension-5 term
\be
{\cal L}_{\rm int} = -\frac{1}{\Lambda} H^\dagger H \bar{\psi} \psi,
\ee
where the new scale $\Lambda$ absorbs the ${\cal O}(1)$ coupling constant.
The dimension-6 term described by the four fermion interaction
$\sim 1/\Lambda^2 (\bar{Q}_L \gamma^\mu Q_L + \bar{q}_R \gamma^\mu q_R)
(\bar{\psi} \gamma_\mu \psi)$ is possible
and contributes to the annihilation cross section 
through $\bar{\psi} + \psi \to f + \bar{f}$. 
However it is suppressed by the factor of $M^2/\Lambda^2$ 
where $M$ is the physical mass of $\psi$
in the most region of the allowed parameter space
and we ignore the dimension-6 term in this work.
Although Eq. (4) is a non-renormalizable term, a renormalizable interaction
is derived from this term after the electroweak symmetry breaking.
With the vacuum expectation value (VEV) of the electroweak symmetry breaking
\be
H \to H + \frac{1}{\sqrt{2}} \left(
\begin{array}{c}
  0 \\
  v \\
\end{array}
\right),
\ee
we rewrite the Lagrangian
\be
{\cal L}_{\rm DM} + {\cal L}_{\rm int}
 = \bar{\psi} ~i \gamma^\mu \partial_\mu \psi -M \bar{\psi} \psi
     - g_{\psi} h \bar{\psi} \psi - \frac{1}{2 \Lambda} h^2 \bar{\psi} \psi,
\ee
where $h$ is the SM Higgs boson and the Yukawa type coupling $g_\psi$ 
and shifted mass $M$ are given by
\be
M = m_0 + \frac{v^2}{2 \Lambda},~~~~g_\psi = \frac{v}{\Lambda},
\ee
with the electroweak scale $v=246$ GeV.
Thus $g_\psi$ is the only bridge between the dark matter sector
and the SM sector.
In this model, we introduce two new parameter $m_0$ and $\Lambda$,
or equivalently the physical mass of the singlet fermion $M$ 
and the coupling $g_\psi$.
While the dimension-5 term 
$-(1/2\Lambda) h^2 \bar{\psi} \psi$ in Eq. (6) contributes
to $ \bar{\psi} + \psi \to h + h$ annihilation process,
this channel is not so important for the parameter space 
considered in this work, $i.e,$ $M < m_h$.

\section{Cosmological Implication}

In the early universe, the dark matter fermion $\psi$ is assumed
to be in thermal equilibrium.
When the temperature of the universe is larger than the DM mass, 
the DM number density is given by its thermal equilibrium density.
Once the temperature drops below the DM mass, the number density
is suppressed exponentially so that DM annihilation rate becomes
smaller than the Hubble expansion rate at a certain point. 
Then the DM number density in a comoving volume remains constant.
Therefore, the cosmological abundance depends upon the annihilation cross
section of $\psi$ into the SM matters.
Since $\psi$ couples only to the Higgs boson,
the annihilation processes are the Higgs-mediated $s$-channel processes
and the dominant final states of the annihilation are
$b \bar{b}$, $W^+ W^-$, $Z Z$ and $t \bar{t}$ depending upon 
the center-of-momentum energy $s$.
We have the cross sections
\be
&&v_{\rm rel} \cdot \sigma(\bar{\psi} \psi \to b \bar{b}) 
  = \frac{G_F N_c g_\psi^2 m_b^2}{4 \sqrt{2} \pi} \beta_b^{\frac{3}{2}}
      \frac{(s-4M^2)}{(s-m_h^2)^2 + m_h^2 \Gamma_h^2},
\nonumber \\
&&v_{\rm rel}  \cdot \sigma(\bar{\psi} \psi \to W^+ W^-) 
  = \frac{G_F g_\psi^2 }{8 \sqrt{2} \pi} s \beta_W^{\frac{1}{2}}
      \left( 1-x_W + \frac{3}{4} x_W^2 \right) 
      \frac{(s-4M^2)}{(s-m_h^2)^2 + m_h^2 \Gamma_h^2},
\nonumber \\
&&v_{\rm rel}  \cdot \sigma(\bar{\psi} \psi \to Z Z) 
  = \frac{G_F g_\psi^2 }{16 \sqrt{2} \pi} s \beta_Z^{\frac{1}{2}}
      \left( 1-x_Z + \frac{3}{4} x_Z^2 \right) 
      \frac{(s-4M^2)}{(s-m_h^2)^2 + m_h^2 \Gamma_h^2},
\ee
where $\beta_X = \sqrt{1-4 m_X^2/s }$.
The pair annihilation cross sections of $\psi$ are
thermally averaged over $s$.
The thermal average is given by
\be
\langle \sigma_{\rm ann.} v_{\rm rel} \rangle
=\frac{1}{8 M^4 T K_2^2(M/T)}
  \int_{4 M^2}^{\infty} ds~ \sigma_{\rm ann.} (s) (s - 4 M^2)
               \sqrt{s} K_1 \left(\frac{\sqrt{s}}{T} \right),
\ee
where $K_{1,2}$ are the modified Bessel functions.

\begin{figure}[ht]
\begin{center}
\hbox to\textwidth{\hss\epsfig{file=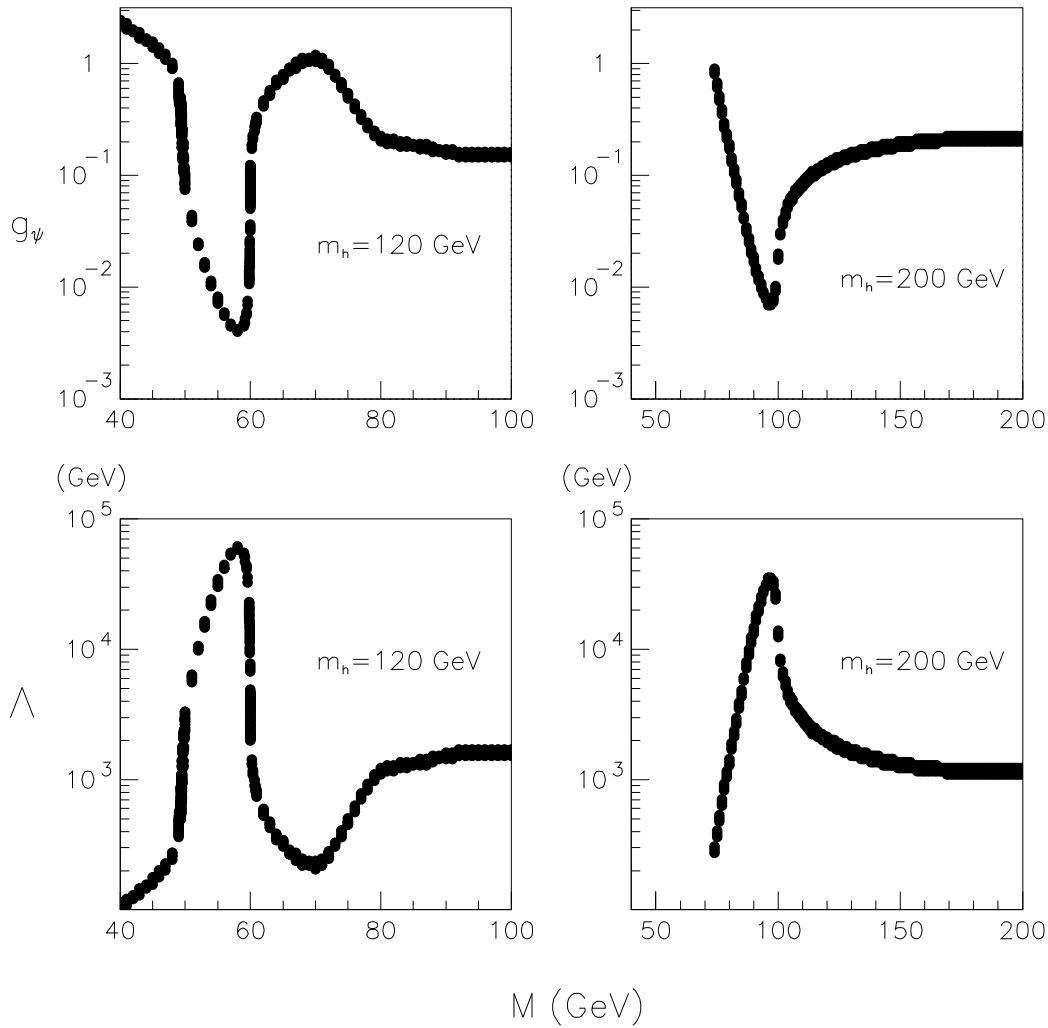,width=16cm}\hss}
\vspace{0.1cm}
\caption{
Allowed parameter set $(M, g_\psi)$ and $(M, \Lambda)$
for Higgs masses $m_h = 120$ and 200 GeV.
}
\end{center}
\end{figure}

The evolution of the relic density is described by the Boltzmann equation
in terms of the $\langle \sigma_{\rm ann.} v_{\rm rel} \rangle$
and the equilibrium number density
\be 
\frac{dn_\psi}{dt} + 3 H n_\psi = -\langle \sigma_{\rm ann.} v_{\rm rel} \rangle
                  \left[ n_\psi^2 -\left( n_\psi^{\rm EQ} \right)^2 \right],
\ee
where $H$ is the Hubble parameter and $n_\psi^{\rm EQ}$ 
the equilibrium number density of $\psi$.
After the freeze out of the annihilation processes,
the actual number of $\psi$ per comoving volume,
$n_\psi/S = n_{\bar{\psi}}/S$ becomes constant
and the present relic density $\rho_\psi = M n_\psi$ is determined.
Approximately we have
\be
\Omega_\psi h^2 \approx \frac{(1.07 \times 10^9) x_F}
                           {\sqrt{g_*}m_{\rm Pl} ({\rm GeV}) 
                  \langle \sigma_{\rm ann.} v_{\rm rel} \rangle},
\ee
where $g_*$ counts the effective degrees of freedom in equilibrium.
The inverse freeze out temperature $x_F \equiv M/T_f$ 
is determined by the iterative equation
\be
x_F = \ln \left( \frac{M}{2 \pi^3} \sqrt{\frac{45 m_{\rm Pl}^2}{2g_*x_F}} 
         \langle \sigma_{\rm ann.} v_{\rm rel} \rangle \right).
\ee
Figure 1 shows the allowed parameter set $(M,g_\psi)$
or equivalently $(M,\Lambda)$ which satisfy the WMAP measurement
of $\Omega_{\rm CDM}h^2$.
The valley in $(M,g_\psi)$ space or the peak in $(M,\Lambda)$ space
indicates the resonant region of the Higgs boson exchange, where $2 M \simeq m_h$.
In that region, the coupling constant $g_\psi$ should be small 
in order to compensate the enhancement of the cross section 
by the Higgs resonance effect.
In the case of $m_h=120$ GeV, one more step appears when $M \sim 80$ GeV.
It denotes that the annihilation channel $\bar{\psi} \psi \to W^- W^+ / ZZ$
opens as $M$ increases.
In the case of $m_h=200$ GeV, the step is hidden by the Higgs resonance effect.
We find that the dark matter with the mass $M > {\cal O}(10 {\rm GeV})$ 
can explain the measured relic density and implies
the new physics with $O(\Lambda) \sim 1$ TeV.

\section{Direct Detection}

We can detect the Weakly Interacting Massive Particle (WIMP) type cold dark matter directly
through its elastic scattering  on target nuclei at underground experiments \cite{witten}.
The elastic cross section for the scattering off a nucleon 
is described by the effective Lagrangian at the hadronic level,
\be
{\cal L}_{\rm eff} = f_p (\bar{\psi} \psi) (\bar{p} p)
               + f_n (\bar{\psi} \psi) (\bar{n} n),
\ee
where the coupling constant $f_p$ is given by \cite{nihei,ellis}
\be
\frac{f_{p,n}}{m_{p,n}} = \sum_{q=u,d,s} f_{Tq}^{(p,n)} \frac{\alpha_q}{m_q}
            + \frac{2}{27} f_{Tg}^{(p,n)} \sum_{q=c,b,t} \frac{\alpha_q}{m_q},
\ee
with the matrix elements 
$m_{(p,n)} f_{Tq}^{(p,n)} \equiv \langle p,n| m_q \bar{q} q |p,n \rangle$
for $q= u,d,s$ and $f_{Tg}^{(p,n)} = 1-\sum_{q=u,d,s} f_{Tq}^{(p,n)}$.
The numerical values of the hadronic matrix elements $f_{Tq}^{(p,n)}$
are determined \cite{ellis}
\be
f_{Tu}^{(p)} = 0.020 \pm 0.004,~~~
f_{Td}^{(p)} = 0.026 \pm 0.005,~~~
f_{Ts}^{(p)} = 0.118 \pm 0.062,
\ee
and
\be
f_{Tu}^{(n)} = 0.014 \pm 0.003,~~~
f_{Td}^{(n)} = 0.036 \pm 0.008,~~~
f_{Ts}^{(n)} = 0.118 \pm 0.062.
\ee
Actually the dominant contribution comes from $f_{Ts}^{(p,n)}$
and therefore we may set $f_p \approx f_n$.
The effective coupling constant $\alpha_q$ is defined
by the spin-independent four-fermion interaction of quarks 
and the dark matter fermion in our model.
The relevant Lagrangian is given by 
\be
{\cal L}_{\rm int} = \sum_q \alpha_q (\bar{\psi} \psi) (\bar{q} q),
\ee
where $\alpha_q$ is obtained from the Higgs exchange 
$t$--channel diagram, such as
\be
\alpha_q = \frac{1}{2} \frac{ g g_\psi}{M^2-m_h^2} \frac{m_q}{m_W}.
\ee

\begin{figure}[ht]
\begin{center}
\hbox to\textwidth{\hss\epsfig{file=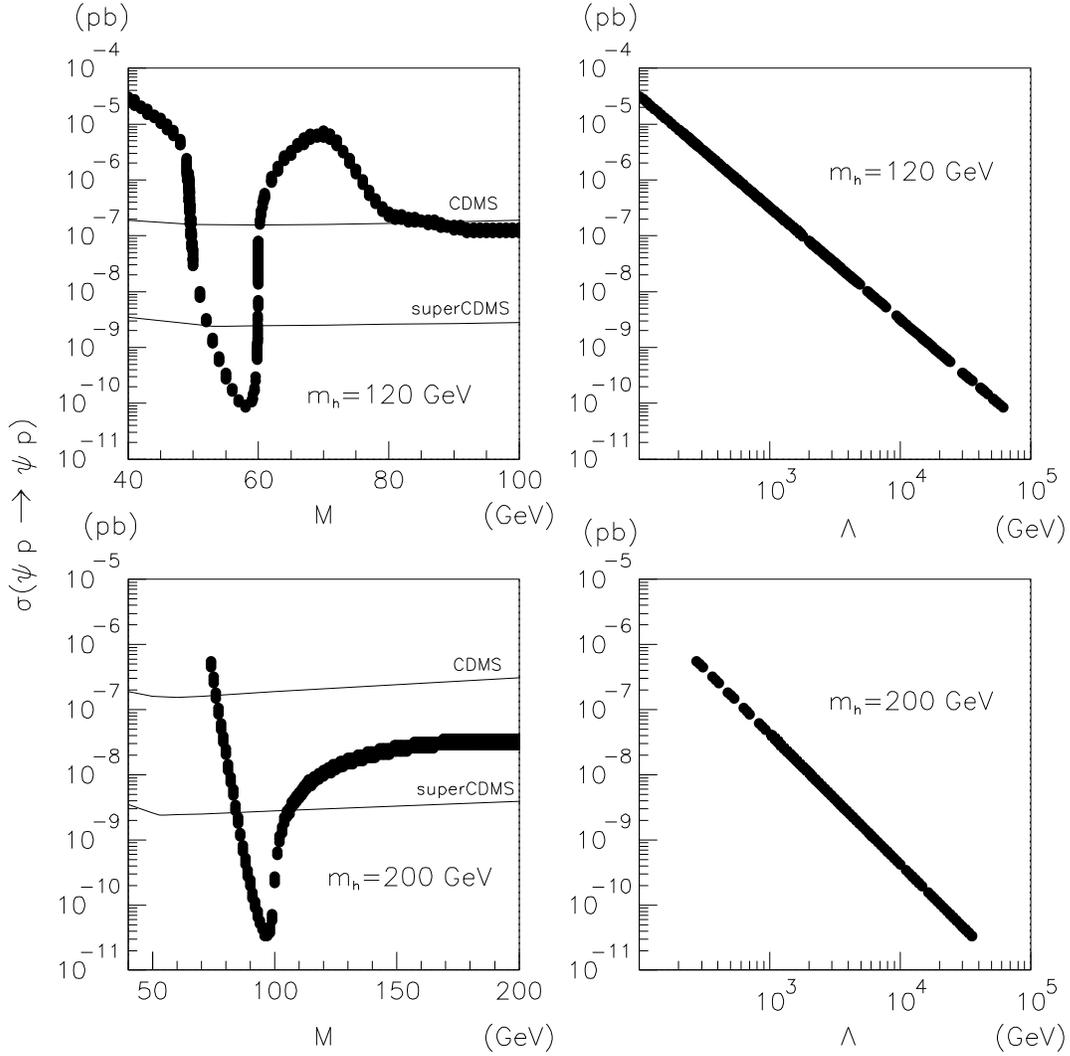,width=16cm}\hss}
\vspace{0.1cm}
\caption{
Predictions of cross section $\sigma(\psi p \to \psi p)$
with respect to $M$ and $\Lambda$
for allowed parameter set $(M, g_\psi)$ or $(M, \Lambda)$.
Higgs masses are assumed to be $m_h = 120$ and 200 GeV.
The thin lines denote the present bounds of CDMS and GENIUS.
}
\end{center}
\end{figure}

We obtain the elastic scattering cross section from the effective Lagrangian Eq. (13), 
which is given by
\be
\sigma = \frac{4 M_r^2}{\pi} \left[ Z f_p + (A-Z) f_n \right]^2
       \approx \frac{4 M_r^2 A^2}{\pi} f_p^2,
\ee
where $M_r$ is the reduced mass of $\psi$ and target defined by
$1/M_r = 1/m_\psi + 1/m_{\rm nuclei}$.
It is convenient to consider the cross section with the single nucleon
for comparing with the experiment. 
Then the target mass is the nucleon mass and 
the cross section is given by
\be
\sigma(\psi p \to \psi p) \approx \frac{4 m_r^2}{\pi} f_p^2,
\ee
where
\be
\frac{1}{m_r} = \frac{1}{m_\psi} + \frac{1}{m_p}.
\ee
The predicted cross sections $\sigma(\psi p \to \psi p)$ 
with the allowed parameter set given in the previous section 
are shown in Fig. 2.
The experimental bound obtained by CDMS collaboration \cite{cdms} 
and the expected reach of superCDMS collaboration \cite{supercdms} 
are also presented.
At present the CDMS bound tells us that the lower bound of the new scale
is ${\cal O}$(TeV).
We find that the future superCDMS experiment can probe 
the most parameter set of the model except for the Higgs resonance region.

\section{Concluding Remarks}

We propose the minimal model of fermionic cold dark matter.
The gauge singlet fermion is introduced 
and coupled to the SM sector through the higher dimensional operator.
The physical interaction is described by the Yukawa type interaction 
after the electroweak symmetry breaking. 
In this model we do not demand the additional symmetry
like $R$ parity nor $Z_2$ parity.
Instead, for the singlet fermion to avoid mixing with 
the SM fermions, we assign the baryon and lepton number 
of the singlet to be zero and require the baryon number and lepton number 
to be preserved.
Only two parameters are involved in our model, one denotes the
new physics scale and the other the mass of the dark matter.
We show that our `minimal' model can explain the current experimental
data on the relic abundance and satisfy the bound of direct detection.
In the future, we expect that the direct search of the superCDMS 
can probe significant portion of the model parameter space.

\acknowledgments
This work is supported by 
the BK21 program of Ministry of Education (K.Y.L.),
and the Astrophysical Research Center for the Structure and Evolution
of the Cosmos funded by the KOSEF (Y.G.K.).


%

\def\PRD #1 #2 #3 {Phys. Rev. D {\bf#1},\ #2 (#3)}
\def\PRL #1 #2 #3 {Phys. Rev. Lett. {\bf#1},\ #2 (#3)}
\def\PLB #1 #2 #3 {Phys. Lett. B {\bf#1},\ #2 (#3)}
\def\NPB #1 #2 #3 {Nucl. Phys. {\bf B#1},\ #2 (#3)}
\def\ZPC #1 #2 #3 {Z. Phys. C {\bf#1},\ #2 (#3)}
\def\EPJ #1 #2 #3 {Euro. Phys. J. C {\bf#1},\ #2 (#3)}
\def\JHEP #1 #2 #3 {JHEP {\bf#1},\ #2 (#3)}
\def\IJMP #1 #2 #3 {Int. J. Mod. Phys. A {\bf#1},\ #2 (#3)}
\def\MPL #1 #2 #3 {Mod. Phys. Lett. A {\bf#1},\ #2 (#3)}
\def\PTP #1 #2 #3 {Prog. Theor. Phys. {\bf#1},\ #2 (#3)}
\def\PR #1 #2 #3 {Phys. Rep. {\bf#1},\ #2 (#3)}
\def\RMP #1 #2 #3 {Rev. Mod. Phys. {\bf#1},\ #2 (#3)}
\def\PRold #1 #2 #3 {Phys. Rev. {\bf#1},\ #2 (#3)}
\def\IBID #1 #2 #3 {{\it ibid.} {\bf#1},\ #2 (#3)}

\end{document}